\documentclass[prl,aps,twocolumn,preprintnumbers,floats,amsmath,amssymb]{revtex4}

\usepackage{graphicx}
\usepackage{dcolumn}
\usepackage{bm}
\usepackage{amssymb}
\usepackage[latin1]{inputenc}

\def \be{\begin{equation}}
\def \ee{\end{equation}}

\def\a{\sigma}

\begin{document}

\title{Do metals exist in two dimensions? 
\\A study of many-body localisation in low density electron gases.}

\author{Geneviève Fleury and Xavier Waintal}
\affiliation{Nanoelectronics group, Service de Physique de l'Etat Condensé,
CEA Saclay F-91191 Gif-sur-Yvette Cedex, France\\}

\date{\today}

\begin{abstract}
Using a combination of ground state quantum Monte-Carlo and finite size scaling techniques,
 we perform a systematic study of the effect of Coulomb interaction on the
localisation length of a disordered two-dimensional electron gas.
We find that correlations delocalise the 2D system. In the absence of valley degeneracy 
(as in GaAs heterostructures), this delocalization effect corresponds to a finite increase of the
localization length. The delocalisation is much more dramatic in the presence of valley degeneracy 
(as in Si MOSFETSs) where the localization length increases drastically. Our results suggest that a rather simple mechanism can account for the
main features of the metallic behaviour observed in high mobility Si MOSFETs. Our findings support the claim that this
behaviour is indeed a genuine effect of the presence of electron-electron interactions, yet that 
the system is not a ``true'' metal in the thermodynamic sense.
\end{abstract}
\maketitle
Since the discovery of Anderson localisation~\cite{anderson1958} 
and the subsequent scaling theory of localisation~\cite{abrahams1979} in 1979, it has 
been believed that any disorder, no matter how weak, would drive a two dimensional (2D) system toward an
 insulating state. This absence of metals in 2D has been challenged by the observation~\cite{kravchenko1994} in 1994 of a Metal-Insulator
 Transition in high mobility Silicon MOSFETs. This first set of experiments has been followed by similar
 observations~\cite{abrahams2001,kravchenko2004} in a wide range of 2D systems. However, as there seemed to be no room for this new metal within the widely accepted theory~\cite{lee1985}, the experimental data remained a puzzle to the community. Many theoretical
scenarios were proposed, ranging from rather extreme~\cite{punnoose2005,anissimova2007,spivak2003} to fairly conservative~\cite{altshuler2000,meir1999}. From a theoretical point of view, the difficulty in dealing with these low density 2D systems is double, as both disorder and interaction must be treated in a non pertubative way. Indeed, despite important
efforts, even the non-interacting problem (Anderson localization in 2D) has resisted all analytical approaches so far. 

Here, we take a different route, similar in spirit to what has been done
numerically~\cite{kramer1993} to verify the assumptions of (non-interacting) scaling theory of localization.
Building on a numerical technique~\cite{fleury2008} that we have developped recently, we calculate the many-body
localisation length of the interacting problem. 

We consider  the many-body generalisation of the Anderson model used to
study localisation. The system (with $N$ particles in a lattice made
of $L_x\times L_y$ sites) is parametrized by two dimensionless numbers
that are denoted traditionally $r_s$ (strength of the Coulomb interactions)
and $1/k_F l$ (strength of the disorder).
 When the electronic density $n_s$ decreases, both interaction $r_s=m^* e^2/(\hbar^2 \epsilon\sqrt{\pi n_s})$ 
and disorder $1/k_F l$ increase in a fixed ratio  $\eta= r_s\sqrt{k_F l}$ which characterizes a given sample
 ($m^*$ effective mass, $e$ electron charge, $\epsilon$ dielectric constant, 
$k_F$ Fermi momentum and $l$ mean free path). 
We distinguish three types of systems. Polarised electrons are effectively spinless and the corresponding system 
is referred as the one component plasma (1-C, $M=1$). Non polarised electrons 
(like electrons in GaAs heterostructures) correspond to the two components 
plasma (2-C, $M=2$). The non polarised system with valley degeneracy (electrons in Si MOSFETs) is the four components plasma (4-C, $M=4$). The Hamiltonian for the $M$ components plasma reads,
\be
\label{eq:model}
H=-t\sum_{\langle\vec r\vec r'\rangle \a}c_{\vec r \a}^\dagger c_{\vec r' \a} + \sum_{\vec r \a}
v_{\vec r}n_{\vec r \a}
+\frac{U}{2} \sum_{\vec r \a\ne\vec r'\a'} V_{\vec r-\vec r'} n_{\vec r \a} n_{\vec r' \a'}
\ee
where $c_{\vec r \a}^\dagger$ et $c_{\vec r \a}$ are the usual creation and annihilation operators of 
one electron on site $\vec r$ with inner degree of freedom $\a$, 
the sum $\sum_{\langle\vec r\vec r'\rangle\a}$ is restricted to nearest 
neighbours and $n_{\vec r \a}=c_{\vec r \a}^\dagger c_{\vec r \a}$ is the density operator. The internal degree
of freedom $\a=1\dots M$ corresponds to the spin ($M=2$ for semi-conductor heterostructures like GaAs/AlGaAs)
or the spin and valley degeneracy ($M=4$ for Si MOSFETs).
The disorder potential $v_{\vec r}$ is uniformly distributed inside $\left[-W/2,W/2\right]$. 
$t$ is the hopping parameter and $U$ is the effective strength of the two body interaction $V_{\vec r}$. To reduce finite size 
effects, $V_{\vec r}$, whose expression can be found in~\cite{waintal2006}, is obtained from the bare Coulomb interaction using the Ewald summation 
technique. We work at small filling 
factor $\nu\equiv N/(L_x L_y)\ll 1$, where we recover the continuum limit. The two dimensionless parameters 
read $r_s=U/(2 t \sqrt{\pi \nu})$ and $k_F l = 192\pi \nu t^2/(M W^2)$.

Without interaction ($r_s=0$), scaling theory of localisation~\cite{abrahams1979} predicts that the scaling 
function $\beta(g)=d\log g/d\log L$ (which indicates how the conductance $g$ of the system changes as one increases
 its size $L$) is a function of  $g$ only. In two dimensions, $\beta(g)$ is always negative  
(for arbitrary disorder $1/k_Fl> 0$) so that $g$ always extrapolates to zero in the thermodynamic limit. 
It was found in Ref.~\cite{fleury2008} (to which we refer for technical details about the numerical method) that, for the 1-C plasma, not only the non-interacting $\beta(g)$ function is unaffected by the interaction, but upon increasing $r_s$, 
the system becomes even more insulating. In this letter, we show that the 2-C and 4-C plasma have
 an opposite behaviour from 1-C and get delocalised by electronic correlations.
\begin{figure}
\includegraphics[width=8.5cm]{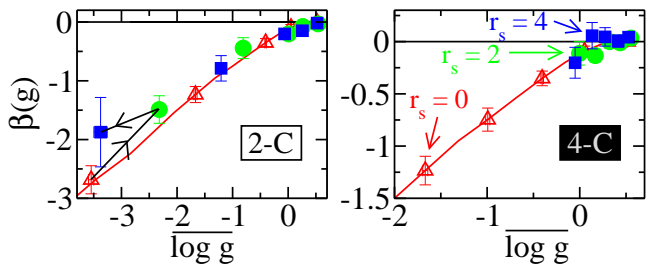}
\caption{\label{fig:beta} 
$\beta(g)$ as a function of $\overline{\log g}$ for the 2-C ($N$=32, left) and 4-C system ($N$=64, right). The red line is the non-interacting scaling function. Symbols stand for
$r_s$=0 (triangles), $r_s$=2 (circles) and $r_s$=4 (squares), for $0.55<k_F l<24.5$. 
The black arrow in the left panel follows a given value of disorder for increasing $r_s$.}
\end{figure}

Let us start with the study of the $\beta (g)$ function, shown in Fig.~\ref{fig:beta}. We do not assume that $\beta$ 
is a function of $g$ only, but we merely plot simultaneously $\beta=d\log g/d\log L$ and $g$ while varying both $k_F l$ and $r_s$.
Upon switching on the interaction in the 2-C system, $g$ and $\beta$ first increase quickly.
They reach a maximum at $r_s\approx 2$ and then decrease. We observe no visible deviation from one-parameter scaling:
even though $g$ and $\beta$ increase with $r_s$, the trajectory $[g(r_s),\beta(r_s)]$ remains on the non-interacting
$\beta(g)$ curve. In particular, at  $r_s\approx 2$ where the delocalisation effect is maximum, the
 value of $\beta$ always remains strictly negative $\beta<0$ so that, even though the localisation length $\xi$
has increased
 significantly, it is still finite, and the system would still be insulating in the thermodynamic limit. 
The situation for the 4-C system is qualitatively different from 2-C as
 the delocalisation effect is much more pronounced: $\beta$ quickly increases with $r_s$ and reaches $\beta=0$ i.e. a
 metallic state up to our statistical precision. 
While the non interacting points are spread over a wide range of $g$ and $\beta$, once the interaction is switched on, 
all the points (but one obtained for the strongest disorder) move toward $\beta=0$. Hence, the localisation length $\xi$ becomes much larger than the system
size and the system becomes practically a metal. Yet, as in 2-C, we observe no visible deviation from one parameter 
scaling so that even though $\xi$ has increased dramatically we cannot conclude that the system becomes 
a ``true'' metal in the thermodynamic sense (i.e. a divergence of $\xi$).
We note that indications that spin and valley degeneracy could play a positive role for transport properties 
were found in the limit of low disorder and interaction ($k_F l \rightarrow \infty$, $r_s\ll 1$) 
as early as 1983~\cite{finkelshtein1983,castellani1984}. However, these works predicted deviations from
one parameter scaling which we did not observe in our (non-perturbative) regime of parameters 
($0.5<k_F l <4$ , $0\le r_s\le 10$).
\begin{figure}
\includegraphics[keepaspectratio,width=8.5cm]{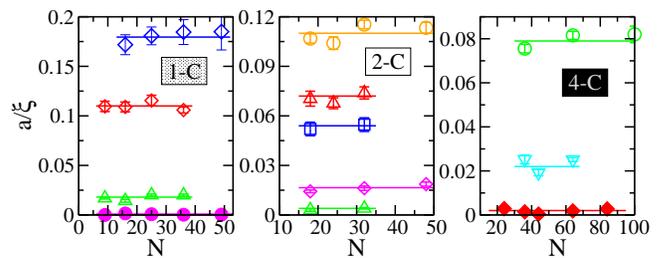}
\caption{\label{fig:xsiN} 
$a/\xi$ as a function of $N$ for 1-C (left), 2-C (middle) and 4-C systems (right panel). 
The symbols correspond to different strengths of interaction ($r_s=0$ (circles), $r_s=0.25$ (down triangles), $r_s=2$ (diamonds), $r_s=4$ (up triangles) and $r_s=6$ (squares)) and of disorder ($1<k_Fl<6$ for 1-C, $1<k_F l<2.4$ for 2-C and $k_Fl=1.05$ for 4-C).}
\end{figure}

As we observed no deviation to one 
parameter scaling, it means that all the relevant information of the system is encapsuled in the
(one parameter) localization length $\xi(r_s,k_Fl)$ of the system.
Integrating the $\beta(g)=d\log g/d\log L$ function, one parameter scaling implies that 
\be
\label{eq:defxi}
\frac{1}{\xi}\equiv \frac{1}{L} \exp\int_{-1.5}^{\log g} \frac{d\log u}{\beta[\log u]}
\ee
is independent of the size of the system. We have carefully tabulated $\beta(g)$ without interaction
and we use Eq.(\ref{eq:defxi}) to extract $\xi$. Note that even if there are deviations from the non-interacting $\beta (g)$, those are small in the regime of parameters that we have considered, 
and can be safely ignored for our purpose. We also emphasize that this method allow us to extract $\xi$ even when $\xi\approx L$ and that we recover $g=g_0 e^{-L/\xi}$ for $L\gg\xi$.
In Fig.~\ref{fig:xsiN}, we plot $a/\xi$ as a function
of $N$ for various values of disorder and interaction ($a\equiv 1/\sqrt{\pi\nu}$ is the averaged distance
between particles). Except for very small $N$
(not shown), we find that $a/\xi$ is indeed independent of $N$ for all three systems 1-C, 2-C and 4-C.
Note that for 4-C, we show only one value of $k_Fl$ for different $r_s=0$,  $r_s=0.25$ and $r_s=2$.
For the latter, $a/\xi=0$ up to our statistical resolution while the corresponding non-interacting 
system is localized. Hence, the observed (strong) delocalization effect
is robust against increasing $N$. 

\begin{figure}
\includegraphics[width=8.5cm]{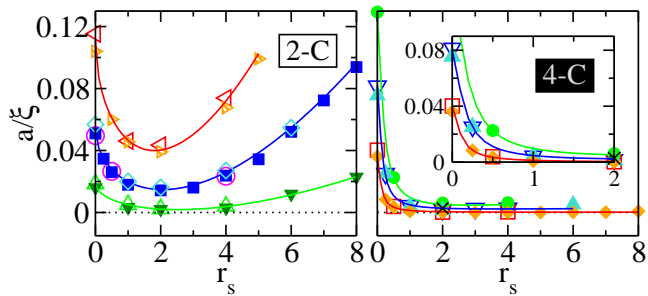}
\caption{\label{fig:xsi} 
Inverse localisation length $a/\xi(r_s)$ for the 
2-C (left) and 4-C systems (right) at various $N$ and $k_F l$. For 2-C: top to bottom: 
$k_F l$=1.05 ($N$=32 left triangles, $N$=24 right triangles), $k_F l$=1.51 
($N$=32 diamonds, $N$=24 circles, $N$=18 squares),  $k_F l$=2.36 ($N$=32 up triangles, $N$=18 down triangles).
For 4-C: $k_F l$=0.77 ($N$=36 circles), 
$k_F l=1.05$ ($N$=84 cross, $N$=64 down triangles, $N$=36 up triangles), 
$k_F l$=1.51 ($N$=64 squares, $N$=36 diamonds). Inset: zoom of the main figure. 
Lines are guides to the eye. Error bars are roughly equal to symbols sizes.}
\end{figure}

The corresponding localization lengths have been collected in Fig.~\ref{fig:xsi} 
which is the chief result of this letter. It is, to the best of our
knowledge, the first calculation of the localisation length in presence of many-body correlations.

In the rest of this letter, we turn to the implications of these results to the
transport properties of 2D gases. A strong debate has taken place in the literature on
the nature of the observed metallic behaviour~\cite{kravchenko1994} 
(i.e. decrease of resistivity as temperature is lowered). 
We will argue that the above findings provide a simple mechanism that accounts for the
main experimental observations. 
The first observation of importance in those low density systems, is that the Fermi energy is 
extremely low, so that the actual typical temperature 
is not much smaller than the Fermi temperature. This is to be contrasted 
with more conventional systems where only a small fraction of the electrons around the Fermi surface 
take part in the transport properties. Hence, the nature of the (highly) excited states and in particular 
their localisation properties is of crucial importance here. 
In a non-interacting system, the localization length varies very quickly with $k_F l$ as electrons
gather kinetic energy, typically exponentially~\cite{lee1985}. 
An immediate consequence is that the excited states, with higher kinetic energy, are {\it less}
localized than the ground state. In particular, the {\it polarized} system is {\it less} localized than the 
(non polarized) ground state. In such a situation, one naturally expects a negative in plane magnetoresistance: 
an in plane magnetic field $H$ (with no orbital but only Zeeman coupling to the electrons) polarizes the system 
hence delocalizes the system whose resistance decreases. Similarly, upon increasing the temperature $T$, the less
localized excited states get populated and the overall resistance decreases. 

Now, we find that this natural order can be reversed in presence of interactions. One ends up in a situation where the ground
state is delocalized while its excited states are still strongly localized. In this situation,
the behaviour of the resistivity upon increasing an in plane magnetic field $H$ or the temperature is reversed
with respect to the non-interacting case: populating the highly localized excited states leads to an increase of
resistivity, hence a positive in plane magnetoresistance as well as an increase of resistance upon increasing temperature
(as in the ``metallic'' phase of the Si MOSFETs). Both effects are expected to take place when
the corresponding Zeeman energy ${\rm g}\mu_B H/2$ and temperature $kT$ are of the order of the 
characteristic energy on which $\xi$ varies, i.e. the energy $E_P(k_Fl, r_s)$ to polarize the system.   

\begin{figure}
\includegraphics[width=8.5cm]{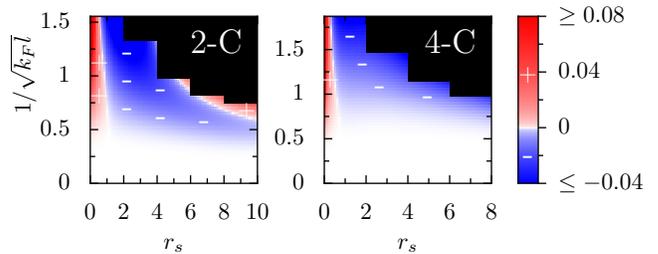}
\caption{\label{fig:phase} 
$\delta(r_s,1/\sqrt{k_Fl})$ for the 2-C (left) and 4-C system (right panel).
$\delta>0$ (red, $+$) 
indicates that the polarized system is less localized than the non-polarized one,
$\delta<0$ (blue, $-$) the opposite. No data are available in the black regions.}
\end{figure}

In the ``phase diagram'' Fig.~\ref{fig:phase}, we have plotted the difference $\delta\equiv a/\xi_{NP} -a/\xi_P$ of the
inverse localization length for the non-polarized $\xi_{NP}$ and polarized $\xi_P$ systems. The polarized system
is an excited state of the non polarized one, and we have verified numerically that the picture that follows also apply to
partially polarized systems (excited states of lower energy). The $(r_s,k_Fl)$
plane can be divided into three regions. In the white region $k_Fl\geq 4$, both localization lengths $\xi_P$ and 
$\xi_{NP}$ are extremely large so that 
$\delta$ is essentially zero. Upon increasing disorder $k_Fl<4$ at small $r_s$, 
the localization length of the system
decreases quickly. Yet, as the interaction is weak, the polarized system is less localized than the
non-polarized one
which leads to a positive $\delta$ (red, $+$). 
In the last region (blue, $-$), the interaction have delocalized
the non-polarized system much more than the polarized one, hence a negative $\delta$. In our view, this (new)
blue region corresponds to the regime where the metallic behaviour has been observed. Note that there is an
important difference between 2-C and 4-C in the blue region: for 4-C, the ground state is
(for practical purposes) delocalized while for 2-C it is not. Hence, for 2-C, temperature has
two conflicting effects: on one hand it activates transport mechanisms such as variable range hopping~\cite{efros1975},
while on the other hand it populates the (more localized) excited states.

Upon varying density, a given sample moves on Fig.~\ref{fig:phase} according
to $r_s=\eta/\sqrt{k_F l}$ (straight lines). In the absence of interaction, 
$\eta= 2\sqrt{\mu} e^{3/2} m^* /(\hbar^{3/2} \epsilon \sqrt{M})$, so that the mobility
$\mu$ obtained at low temperature and large density (small interaction) can be used
as an estimate of $\eta$. We note however that upon varying density, $\eta$ remains constant only
for white disorder. Real disorder is correlated so that $\eta$ is expected to increase at low
density as larger scales are probed. Also, the disorder seen by the electrons at the Fermi level  
is partially screened by the rest so that the above formula overestimates $\eta$.
Rough estimates are $\eta \approx 10$ (high mobility Si MOSFET, $\mu\ge 10^4 {\rm cm^2.V^{-1}.s^{-1}}$), 
$\eta \approx 50$ (electrons in GaAs/GaAlAs heterostructures, $\mu\ge 10^6 {\rm cm^2.V^{-1}.s^{-1}}$),
and $\eta \approx 200$ (holes in GaAs/GaAlAs heterostructures, $\mu\approx 10^6 {\rm cm^2.V^{-1}.s^{-1}}$) so
that high mobility Si MOSFETs are dirtier than their III-V counterparts.

For definiteness, let us concentrate on the data presented in Fig.1 of Ref.~\cite{altshuler2000} 
(hereafter referred as  Fig.A) for a typical high mobility Si MOSFET.
At very high density ($k_Fl\gg 1$, $r_s\ll 1$), the system can be described within diagrammatic
theory which predicts small quantum corrections to transport properties~\cite{lee1985}. This is
the left part of the white region in Fig.~\ref{fig:phase} where one observes, for instance,
the usual signatures of weak localization. The resistivity $\rho(T)$ as a function of temperature
is roughly flat in this region (except at large temperature where phonons set in). 
Upon decreasing the density, one enters the blue region 
of  Fig.~\ref{fig:phase} at 
$k_F l \approx 4$ ($n_s\approx 2.5\ 10^{11} {\rm cm^{-2}}$ for the sample of Fig.~A).
In this region, one observes the strong positive variation of $\rho(T)$ which has been puzzling the 
community for more than a decade. We attribute this behaviour to the fact that in this region, the
ground state is (for all practical purposes) delocalized (see right panel of Fig.~\ref{fig:xsi}) 
while its excitations are not. Hence
upon increasing temperature those localized states get populated and the resistivity increases strongly.
This scenario implies in particular that (i) The corresponding characteristic energy scale
where the increase of $\rho(T)$ is expected is the energy $E_P(k_Fl, r_s)$ to polarize the system.
We have calculated numerically this energy scale in the relevant regime and found $E_P\approx 0.15 E_F$ where $E_F$ 
is the Fermi energy. Indeed, one observes in Fig.~A that $T/T_F\approx 0.15$ corresponds to the
crossover temperature where the strong increase of $\rho(T)$ takes place.
(ii) An in plane magnetic field polarizes the system which becomes localized, 
hence a positive magnetoresistance is expected with a variation of resistance of the same order of
magnitude as the variation due to the change in temperature. This is indeed observed for magnetic fields $H$
that correspond to a Zeeman energy of the order of $E_P$ (i.e. $\approx 1.6 T$ at $n_s=10^{11}{\rm cm^{-2}}$)
as can be seen, for instance, in Fig.10 of Ref.\cite{abrahams2001}. 

To conclude, we have reported on the first quantitative calculation of the localization length
in an interacting two dimensional electronic gas. The scenario that naturally emerges from the data captures
all the essential features of the experiments at a semi-quantitative level, with no adjustable parameters. 
In particular, our data account for: 
(i) why the metallic behaviour is found at low
density, i.e. when the corresponding polarized system lies in the vicinity of the quantum of conductance.
(ii) Why it is destroyed by an in plane magnetic field. 
(iii) Why Si MOSFETs, with valley degeneracy,
show a much stronger metallic behaviour than other materials. 
(iv) Why disorder matters, and samples with $\eta<1$
will not exhibit the metallic behaviour (red part of Fig.~\ref{fig:phase}) while too clean samples (white part)
will only show a weak signal. 
(v) The characteristic magnetic field and temperature for the dependance of the
resistivity.

We have not discussed the metal-insulator transition {\it per se}, but rather focused on the existence
of the metallic behaviour. Indeed, for large interaction $r_s$ and disorder $1/k_F l$, there is no question that the system
will eventually become insulating. Although the nature of this insulator and of the corresponding transition is clearly
an interesting topic, the real mystery lies in the nature of the metal itself. Our results suggest that
this metallic behaviour is due to a non-perturbative interaction delocalization effect. On the other hand, 
we found no significant deviation from the non-interacting one-parameter scaling which suggests that
the 4-C systems are not ``true'' thermodynamic metals and would become insulating at ultrasmall temperature 
(2-C systems are unambiguously insulators).
In any case, we find that the question of the existence of a ``true'' metal is not relevant in 
those experiments as a different physics (of much higher energy) has been probed in practice.

\acknowledgments
{\it Acknowledgements.} Strong support from the CEA supercomputing facilities CCRT is gratefully acknowledged.
We thank V. Rychkov and F. Portier for useful comments on the manuscript.
\bibliography{loc4C}

\end{document}